\documentclass[10pt,twocolumn,prl,superscriptaddress,showpacs,byrevtex]{revtex4}
\usepackage{graphics}
\usepackage{amssymb}

\newcommand{\rh}{\tilde{h}}
\newcommand{\ms}{\tilde{s}}
\newcommand{\umin}{{\rm min}}
\newcommand{\umax}{{\rm max}}

\begin{document}

\title{Nonconcave entropies from generalized canonical ensembles}

\author{Marius Costeniuc}
\email{marius.costeniuc@gmail.com}
\address{Max Planck Institute for Mathematics in the Sciences, Inselstr. 22-26,          
04103 Leipzig, Germany}

\author{Richard S. Ellis}
\email{rsellis@math.umass.edu}
\address{Department of Mathematics and Statistics, University of Massachusetts, Amherst, MA 01003, USA }

\author{Hugo Touchette}
\email{htouchet@alum.mit.edu}
\address{School of Mathematical Sciences, Queen Mary, University of London, London E1 4NS, UK}

\date{\today}

\begin{abstract}
It is well-known that the entropy of the microcanonical ensemble cannot be calculated as the Legendre transform of the canonical free energy when the entropy is nonconcave. To circumvent this problem, a generalization of the canonical ensemble which allows for the calculation of nonconcave entropies was recently proposed. Here, we study the mean-field Curie-Weiss-Potts spin model and show, by direct calculations, that the nonconcave entropy of this model can be obtained by using a specific instance of the generalized canonical ensemble known as the Gaussian ensemble. 
\end{abstract}

\maketitle

Systems involving long-range interactions, such as gravitational forces or wave-particle interactions, have many unusual properties at equilibrium which set these systems apart from those involving short-range interactions \cite{dauxois2002,gross1997}. The root of most, if not all, of these unusual properties can be traced back to the fact that the microcanonical entropy function of long-range systems can be nonconcave as a function of the energy, even in the thermodynamic limit. By constrast, short-range systems can have nonconcave entropies at finite volume, but any nonconcave parts of the entropy must and will disappear in the thermodynamic limit, leaving a thermodynamic entropy which is always concave \cite{ruelle1969}. This has profound physical consequences because the concavity of the entropy determines ultimately whether the microcanonical ensemble (ME) is equivalent with the canonical ensemble (CE). If the entropy is concave, then the two ensembles are equivalent; otherwise they are nonequivalent at both the thermodynamic and macrostate levels \cite{eyink1993,ellis2000}. Thus short-range systems always have equivalent ensembles, but long-range systems need not; they can have equilibrium properties within the ME that have no counterparts in the CE. Many illustrations of this possibility have been given recently for a variety of systems, including those arising in the study of gravitation \cite{lynden1968,chavanis2002}, geostrophic turbulence \cite{ellis2000,ellis2002}, plasmas \cite{smith1990}, and magnetism \cite{ispolatov2000,barre2001,ellis2004,costeniuc2005}.

Systems with nonconcave entropies are special not only in terms of their equilibrium properties, but also in the way in which their entropy can be calculated. In cases of ensemble equivalence, it is well known that the microcanonical entropy function can be calculated as the Legendre transform of the canonical free energy function.
This way of calculating the entropy goes back to Gibbs, and is nowadays the way of choice for calculating entropies because of the many practical advantages it offers. 
One well known advantage is that calculations carried out at the level of the free energy are generally more tractable than those carried out at the level of the entropy directly. Another advantage is that the definition of the free energy allows for many approximation schemes (e.g., perturbative expansions or variational principles), which are not available in the microcanonical ensemble.

The problem for systems having a nonconcave entropy is that the entropy does not correspond to the Legendre transform of the free energy anymore. This is obvious if one recalls that Legendre transforms yield only concave functions. Therefore, if one knows or suspects that a given system has a nonconcave entropy, then one also knows that this entropy cannot be calculated in the CE as the Legendre transform of the free energy. In this case, one must rely on the ME to perform that calculation (see, e.g., \cite{ellis2004,barre2005}).

Our goal in this paper is to illustrate an alternative method for calculating entropy functions which goes beyond the CE in that it can be used to obtain nonconcave entropies. The method is not based on the ME. Rather, it is based on a \textit{generalized canonical ensemble} (GCE) put forward recently by us, and works by modifying the structure of the Legendre transform through the use of a generalization of the free energy function. We have already presented the theory of the GCE and of its equivalence with the ME in \cite{costeniuc22005,costeniuc2006}; our goal here is to illustrate this theory with the help of a simple spin model known as the mean-field Curie-Weiss-Potts model \cite{ispolatov2000,costeniuc2005}. For this model, we shall calculate a nonconcave entropy function using the GCE, and show that the result agrees, in some appropriate limit, with the one obtained in the ME. In doing so, we shall explain a number of practical aspects of the GCE which have not been fully discussed before. 

To begin, we review the definition of the GCE and the results establishing equivalence of this ensemble with the ME. Following our previous work \cite{costeniuc22005,costeniuc2006}, we define the GCE by the following probability density over the microstates $\omega $: 
\begin{equation}
P_{n,g,\alpha }(\omega )=\frac{e^{-\alpha nh(\omega )-ng(h(\omega ))}}
{Z_{n,g}(\alpha )}.
\end{equation}
In this expression, $h(\omega )=H(\omega )/n$ is the energy per particle or mean energy, $\alpha$ is a real parameter, $g(h)$ is a continuous but otherwise arbitrary function of $h$, and $Z_{n,g}(\alpha )$ is a normalization constant given by
\begin{equation}
Z_{n,g}(\alpha )=\int e^{-\alpha nh(\omega )-ng(h(\omega ))}d\omega .
\label{genpart1}
\end{equation}
This function is naturally interpreted as a generalized partition function; from it, we define a generalized free energy via the usual limit 
\begin{equation}
\varphi _g(\alpha )=-\lim_{n\rightarrow\infty} \frac 1n\ln Z_{n,g}(\alpha ).
\label{genfree1}
\end{equation}
It is obvious that $P_{n,g,\alpha }(\omega )$ and, consequently, $Z_{n,g}(\alpha )$ and $\varphi _g(\alpha )$ reduce to their standard canonical-ensemble expressions for $g=0$ \cite{note1}. This explains why we use the term ``generalized'' CE. 

To complete this short review of the GCE, we state the generalized Legendre transform that will be used hereafter to obtain the entropy of the ME from the free energy of the GCE. At this point, we recall the thermodynamic-limit definition of the microcanonical entropy:
\begin{equation}
s(u)=\lim_{n\rightarrow\infty} \frac{1}{n} \ln \int \delta(h(\omega)-u)\ d\omega .
\end{equation}
In terms of $\varphi_g(\alpha)$ we then have the following: 
\textit{If, for a given choice of $g$, $\varphi _g(\alpha )$ is differentiable at $\alpha$, then} 
\begin{equation}
s(u_{g,\alpha })=\alpha u_{g,\alpha }+g(u_{g,\alpha })-\varphi _g(\alpha ),
\label{lf3}
\end{equation}
\textit{where $u_{g,\alpha }=\varphi _g^{\prime }(\alpha )$} (see \cite{costeniuc2006}, Theorem 4).
The choice $g=0$ yields, as expected, the standard Legendre transform
\begin{equation}
s(u_\beta)=\beta u_\beta-\varphi(\beta ),\qquad u_\beta=\varphi'(\beta),
\label{lf4}
\end{equation}
where $\varphi(\beta)=\varphi_{g=0}(\alpha=\beta)$ stands for the free energy of the CE. For the remaining, it is useful to note that $u_{g,\alpha }$ represents the equilibrium mean energy of the GCE with function $g$ and parameter $\alpha $ \cite{touchette2006}. This parallels the case of the CE for which $u_\beta=\varphi'(\beta)$ represents the equilibrium mean energy of the CE with inverse temperature $\beta$.

We now come to the main point of this paper, which is to consider a system known to have a nonconcave entropy, and show that the system's entropy can be derived from the sole point of view of the GCE by first calculating $\varphi_g(\alpha)$ for that system and then by applying to $\varphi_g(\alpha)$ the generalized Legendre transform shown in (\ref{lf3}). The model that we consider for this purpose is the $q$-state mean-field Curie-Weiss-Potts (CWP) model \cite{ispolatov2000,costeniuc2005,wu1982} defined by the Hamiltonian 
\begin{equation}
H(\omega )=-\frac 1{2n}\sum_{i,k=1}^n\delta (\omega _i,\omega _k).
\end{equation}
In this expression, $\delta (x,y)$ is the Kronecker symbol, $\omega_i$ is a spin variable taking values in the set $\Lambda =\{\theta ^1,\theta^2,\ldots,\theta ^q\}$, and $\omega $ represents the complete configuration of $n$ spins, i.e., $\omega =(\omega _1,\omega _2,\ldots ,\omega _n)$. For simplicity, we consider the case $q=3$, so that spins can only take one of three possible values: $\theta ^1$, $\theta^2$ and $\theta^3$. 

Despite the rather simple nature of the CWP model, a complete, analytical calculation of its generalized free energy seems out of reach. For a start, the integral defining the generalized partition $Z_{n,g}(\alpha)$ does not seem to be explicitly solvable, so that the calculation of $\varphi_g(\alpha)$ cannot proceed from a direct evaluation of Eqs.~(\ref{genpart1}) and (\ref{genfree1}). Fortunately, there is an alternative way to calculate $\varphi_g(\alpha)$ suggested by large deviation techniques. It involves three steps \cite{ellis2000,costeniuc22005}:
 
(i) Rewrite the energy per spin $h(\omega )=H(\omega )/n$ as a function of some macrostate or order parameter $\nu(\omega)$. In our case, we choose the macrostate to be the empirical vector $\nu=(\nu_1,\nu_2,\nu_3)$, the $j$th component of which is defined by
\begin{equation}
\nu_j=\frac 1n\sum_{k=1}^n\delta (\omega _k,\theta ^j).
\end{equation}
A short calculation shows indeed that $h(\omega)=\rh (\nu(\omega))$, where $\rh (\nu)=-\frac{1}{2}\langle \nu,\nu \rangle$. The component  $\nu_j$ represents the relative number of spins in $\omega$ equal to $\theta^j$; $0\leq\nu_j\leq1$, $\sum_j \nu_j =1$. For this reason, $\nu(\omega)$ is often called the statistical distribution of spin state or \textit{spin distribution} for short. The function $\rh$ is called the energy representation function.

(ii) Derive the expression of an entropy function $\ms(\nu)$ for the macrostate $\nu$. In the case of the empirical vector, that entropy function is well known to be given by the statistical (Boltzmann-Shannon) entropy
\begin{equation}
\ms(\nu)=- \sum_{j=1}^3 \nu_j \ln \nu_j .
\end{equation}

(iii) Calculate $\varphi_g(\alpha)$, finally, using the following representation formula:
\begin{equation}
\varphi _g (\alpha )=\inf_{\nu}\left\{ \alpha \rh(\nu)+
g(\rh(\nu))-\ms(\nu)\right\}.
\label{repf1}
\end{equation}
The infimum is evaluated over all allowed values of $\nu$, i.e., all triplets $\nu=(\nu_1,\nu_2,\nu_3)$ such that $0\leq\nu_j\leq1$ and $\sum_j \nu_j =1$.

\begin{figure}[t]
\includegraphics{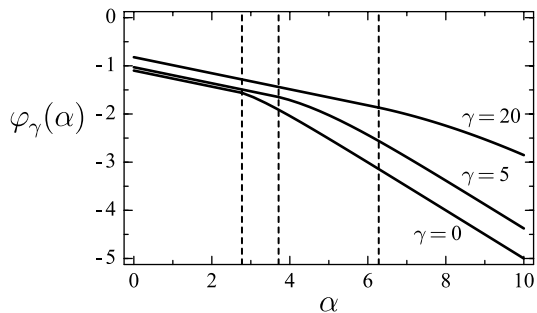}
\caption{Gaussian free energy of the CWP model for different values of $\protect\gamma$.}
\label{figpfree1}
\end{figure}

\begin{figure}[t]
\includegraphics{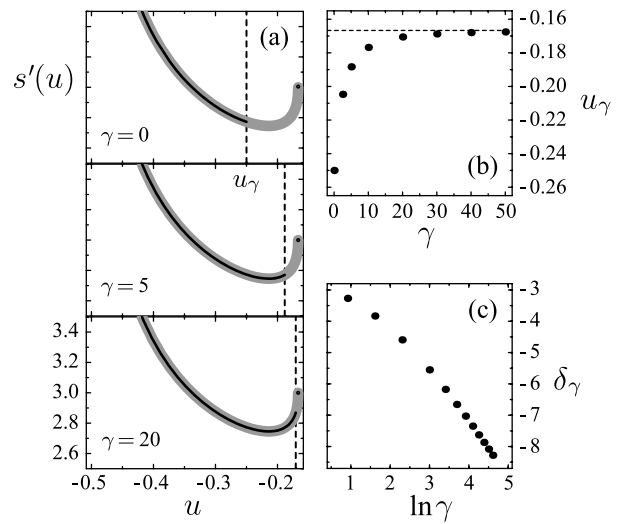}
\caption{(a) Derivative of the entropy function found for the CWP model using the GE (black line) and ME (gray line). The dashed lines show the mean energy value $u_{\protect\gamma}$ up to which there is equivalence between the GE and ME. (b) Convergence of $u_\gamma$ towards $u_\umax$. (c) $\delta_\gamma =\ln | u_\umax - u_\gamma|$ versus $\ln \gamma$.} 
\label{figpent1}
\end{figure}

Equation (\ref{repf1}) was solved numerically using a quadratic function $g$ of the form 
$g(u)=\gamma u^2/2$, where $\gamma\geq0$ \cite{note25}. This choice of $g$ defines a GCE known as the \textit{Gaussian ensemble} (GE) \cite{challa1988}. Fig.~\ref{figpfree1} shows the result of the numerical computation for three increasing values of the parameter $\gamma$. For each value, the free energy of the GE, which we denote by $\varphi_\gamma(\alpha)$, has the particularity that it possesses one nondifferentiable point, located at $\alpha_{\gamma}$, separating two differentiable branches. The left branch is a linear function of $\alpha$ with a slope independent of $\gamma$. The value of the slope is equal to $-1/6$, which is the maximum admissible value of $h$ \cite{note3}. To obtain the value of $s(u)$ at this specific value of $h$, which we denote by $u_\umax$, we simply need at this point to apply Eq.~(\ref{lf3}) to the left branch of $\varphi_\gamma(\alpha)$. The result obtained is $s(u_\umax)=\ln3$, which is the correct value found in the ME corresponding physically to the entropy of the excited state of the CWP model. The right, differentiable branch of $\varphi_\gamma(\alpha)$ can be treated similarly with the difference that $\varphi'_\gamma(\alpha)$ is not a constant but takes values in a range of the form $[u_\umin ,u_\gamma]$, where $u_\umin = \varphi'_\gamma(\infty)=-1/2$ and $u_\gamma= \varphi'_\gamma (\alpha_{\gamma}+0)$. The calculation of Eq.~(\ref{lf3}) for the right branch of $\varphi_\gamma$ therefore yields $s(u)$ in that range. The result is displayed in Fig.~\ref{figpent1}(a) together with the exact entropy function $s(u)$ derived directly from the ME (see Eq.~(4.3) in \cite{costeniuc2005}). Note that, in order to accentuate the relatively shallow nonconcave region of $s(u)$ near $u_\umax$, we have plotted the derivative of the entropy in Fig.~\ref{figpent1}(a) rather than the entropy itself.

The comparison in Fig.~\ref{figpent1}(a) of the entropy obtained in the GE with the ``true'' entropy of the ME shows that the use of the GE enables us to obtain $s(u)$ in a gradual manner by increasing the value of the parameter $\gamma$. For $\gamma=0$, $s(u)$ is recovered from the ground-state mean energy $u_\umin=-1/2$ up to the value $u_{\gamma=0}=-1/4$, while for $\gamma=5$, $s(u)$ is recovered from $u_\umin$ to $u_5\approx-0.1883>u_0$. Increasing the parameter $\gamma$ further to $\gamma=20$, we obtain $s(u)$ from $u_\umin$ up to $u_{20}\approx-0.1706>u_5$. The case $\gamma=0$ corresponds to the CE, so that the part of $s(u)$ obtained in this ensemble corresponds to the concave part of $s(u)$ determined by Maxwell's construction or, equivalently, by the set of supporting lines of that function; see \cite{costeniuc2006} for details. Thus, we see that a virtue of the GE with $\gamma>0$ over the CE is that the former ensemble is able to recover nonconcave points of the entropy function while the latter is not.

By continuing to increase $\gamma$, we can calculate the value $u_\gamma$ up to which the GE is able in to recover $s(u)$. The results are shown in Fig.~\ref{figpent1}(b), as well as in Fig.~\ref{figpent1}(c). The conclusion that we reach from these two figures is that the GE recovers the complete entropy function of the CWP model in the limit $\gamma\rightarrow\infty$, since $u_\gamma\rightarrow u_\umax=-1/6$ in that limit. Hence, for a finite, positive value of $\gamma$, the GE recovers only a part of $s(u)$, but the part of $s(u)$ ``missed'' by the GE can be made arbitrary small by choosing large enough values of $\gamma$. In this sense, we say that the GE and the ME are \textit{asymptotically equivalent}. In general, we know that the GE should be equivalent with the ME at all values of $h$ whenever $\gamma$ is greater than the maximum value of $s''(u)$ (see \cite{costeniuc22005}, Theorem 5.2). In the case of the CWP, $s''(u)$ diverges at $u_\umax$, which explains why the two ensembles become equivalent for all $u\in[u_\umin,u_\umax]$ only in the limit $\gamma\rightarrow\infty$. An analytical study of this limit is presented in \cite{costeniuc22006} (see also \cite{costeniuc22005,costeniuc2006}).

\begin{figure}
\includegraphics{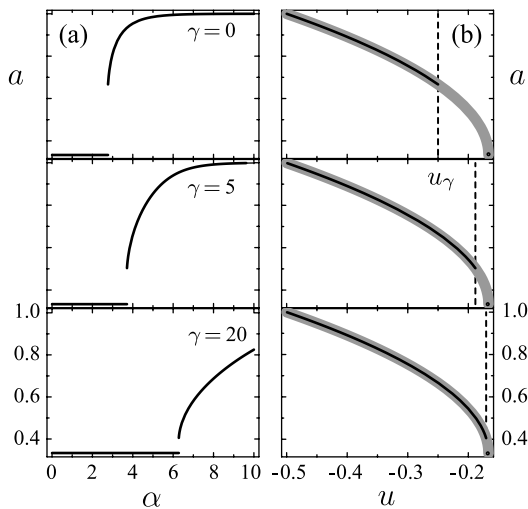}
\caption{(a) First component $a$ of the equilibrium value of $\nu$ in the GE as a function of $\alpha$. (b) Equilibrium value of $a$ in the GE as a function of the equilibrium mean energy of that ensemble (black line); equilibrium value of $a$ in the ME (gray line). 
The dashed lines show the mean energy value $u_{\protect\gamma}$ up to which there is equivalence between the GE and ME.}
\label{figaeq1}
\end{figure}

We can go further in our study of the CWP model by showing that the GE can be used not only to calculate $s(u)$ but also to calculate the microcanonical equilibrium values of $\nu$ of that model. This aspect of the GE, which we refer to as the \textit{macrostate level} of ensemble equivalence, is illustrated in Fig.~\ref{figaeq1}. This figure shows two sets of plots. The one on the left shows the equilibrium values of $\nu$ calculated in the GE as a function of $\alpha$ \cite{note5}, while the one on the right displays the same points but now as a function of their mean energy $u=\rh(\nu)$ (black line). Note that only the first component of $\nu$ is plotted because the equilibrium value of $\nu$ in the GE has the form $\nu=(a,b,b)$ \cite{note25}. The comparison with the ME is established by plotting the equilibrium value of $a$ in the ME (see Eq.~(4.1) in \cite{costeniuc2005}) for all $u\in[u_\umin,u_\umax]$ (gray line).

The results obtained from these calculations show that the GE recovers the equilibrium macrostates of the ME for all $u\in[u_\umin, u_\gamma]$, with $u_\gamma$ approaching $u_\umax$ for increasing $\gamma$. The value $u_\gamma$ here is the same as the one found previously when calculating $s(u)$, so that the range of mean-energy values for which we have equivalence of ensembles at the level of $\nu$ is exactly the range over which we have equivalence of ensembles at the \textit{thermodynamic level}, i.e., the level of $s(u)$. This is not a coincidence but a direct result of the fact that equivalence of the GE and ME at the thermodynamic level entails, essentially, the equivalence of these ensembles at the macrostate level \cite{costeniuc22005,costeniuc2006}. As a result, we can conclude that the GE recovers, in the limit $\gamma\rightarrow\infty$, the microcanonical equilibrium values of $\nu$ for all $u\in[u_\umin,u_\umax]$, since it completely recovers $s(u)$ in the same limit. 

In the end, it is interesting to note that the asymptotic equivalence of the GE and ME is reflected, at the macrostate level, by the fact that the jump of $a$, seen in the GE as a function of $\alpha$, disappears as $\gamma\rightarrow\infty$ (Fig.~\ref{figaeq1}). It can be proved in general that the disappearance of this jump, which is responsible for the nondifferentiable point of $\varphi_\gamma(\alpha)$ (Fig.~\ref{figpfree1}), is a sufficient condition for the complete equivalence of the GE and ME (see \cite{costeniuc2006}, Theorem 4). From a physical point of view, this means that the absence of a first-order phase transition in the GE is a sufficient condition for the equivalence of the GE and ME. 

To summarize, we have considered a mean-field version of the Potts model to show that the nonconcave entropy function of this spin model can be calculated using a generalization of the canonical ensemble known as the Gaussian ensemble. The large deviation formalism that we have used to study this model is totally general in that it can be applied to other models known to have nonconcave entropies. Our future work will aim at studying a number of these models (see, e.g., \cite{touchette22006}), in addition to studying other types of generalized canonical ensembles, including the one defined by the function $g(u)=\gamma |u|$. Although the Gaussian ensemble is thought to be universal, in the sense that it should be able to recover any form of entropy function \cite{costeniuc22005,costeniuc2006}, other generalized ensembles could be useful, in that they may lead to more tractable calculations than those carried out in the Gaussian ensemble (see, e.g., \cite{toral2006}). 

\begin{acknowledgments}
The work of R.S.E. and M.C. was supported by NSF (NSF-DMS-0202309). The work of H.T. was supported by NSERC (Canada) and the Royal Society of London.
\end{acknowledgments}


\begin{thebibliography}{99}
\bibitem{note1} We use $\protect\alpha$ rather than $\protect\beta$ to emphasize the fact that $\protect\alpha$ is not the physical inverse temperature when $g\neq 0$.
\end{thebibliography}
\end{document}